\documentclass{article}%
\usepackage{amsmath}
\usepackage{amsfonts}
\usepackage{amssymb}
\usepackage{graphicx}%
\providecommand{\U}[1]{\protect\rule{.1in}{.1in}}
\newtheorem{theorem}{Theorem}

\newtheorem{remark}[theorem]{Remark}

\newenvironment{proof}[1][Proof]{\noindent\textbf{#1.} }{\ \rule{0.5em}{0.5em}}
\begin{document}

\title{On Equivalent Expressions for the Faraday's Law of Induction}
\author{Fabio G. Rodrigues\\Institute of Mathematics, Statistics and Scientific Computation\\Imecc-Unicamp \\13083-859 Campinas SP, Brazil\\e-mail: fabrod@ime.unicamp.br}
\maketitle

\begin{abstract}
In this paper we give a rigorous proof of the equivalence of some different
forms of Faraday's law of induction clarifying some misconceptions on the
subject \ and emphasizing that many derivations of this law appearing in
textbooks and papers are only valid under very special circunstances and not
satisfactory under a mathematical point of view.

\end{abstract}

\section{Introduction}

Let $\mathbf{\Gamma}_{t}$ be a smooth closed curve in $\mathbb{R}^{3}$ with
parametrization $\mathbf{x(}t\mathbf{,}\ell\mathbf{)}$ which is here supposed
to represent a filamentary closed circuit which is moving in an a convex and
simply-connected (open) region $U\subset\mathbb{R}^{3}$ where at time $t$ as
measured in an inertial frame\footnote{For a mathematical defintion of an
inertial reference frame in Minkowski spacetime see, e.g.,
\cite{rodcap2007,sawu}.}, there are an electric and a magnetic fields
(described after a space orientation is fixed by) $\mathbf{E}:\mathbb{R}%
\times\mathbb{R}^{3}\rightarrow\mathbb{R}^{3},$ $(t,\mathbf{x})\mapsto
\mathbf{E}(t,\mathbf{x})\in\mathbb{R}^{3}$ and$\ \mathbf{B}:\mathbb{R}%
\times\mathbb{R}^{3}\rightarrow\mathbb{R}^{3}$, $(t,\mathbf{x})\mapsto
\mathbf{B}(t,\mathbf{x})\in\mathbb{R}^{3}$. We\ suppose that when in motion
the closed circuit may be eventually \textit{deforming.\ }Let\textit{
}$\mathbf{\Gamma}$ be a smooth closed curve in $\mathbb{R}^{3}$ with
parametrization $\mathbf{x}(\ell)$ representing the filamentary circuit at
$t=0$. Then, the smooth curve $\mathbf{\Gamma}_{t}$ is given by
$\mathbf{\Gamma}_{t}=\mathbf{\sigma}_{t}(\mathbf{\Gamma})$ where
$\mathbf{\sigma}_{t}$ (see details below) is the \textit{flow} of a velocity
vector field $\mathbf{v:}$ $\mathbb{R}\times\mathbb{R}^{3}\mapsto
\mathbb{R}^{3}$, which describes the motion (and deformation) of the closed
circuit. It is an empirical fact known as Faraday's law of induction that on
the closed loop $\mathbf{\Gamma}_{t}$ acts an induced \textit{electromotive
force, }$\mathcal{E}$, such that%
\begin{equation}
\mathcal{E=-}\frac{d}{dt}%
{\displaystyle\int\nolimits_{\mathcal{S}_{t}}}
\mathbf{B}\cdot\mathbf{n}\text{ }da,\label{01}%
\end{equation}
where $\mathcal{S}_{t}$ is a smooth surface on $\mathbb{R}^{3}$ such that
$\mathbf{\Gamma}_{t}$ is its boundary and $\mathbf{n}$ is the normal vector
field on $\mathcal{S}_{t}$. We write $\mathbf{\Gamma}_{t}=\partial
\mathcal{S}_{t}$ with $\mathbf{\Gamma}=\partial\mathcal{S}$. Now, on each
element of $\mathbf{\Gamma}_{t}$ the force acting on a unit charge which is
moving with velocity\ $\mathbf{v}(t,\mathbf{x}(t,\ell\mathbf{))}$ is given by
the Lorentz force law. Thus\footnote{In this paper we use a system of units
such that the numerical value of the speed of light is $c=1$.} the
$\mathcal{E}$ is by definition:%

\begin{equation}
\mathcal{E=}%
{\displaystyle\int\nolimits_{\mathbf{\Gamma}_{t}}}
(\mathbf{E+v\times B)}\cdot d\mathbf{l},\label{02}%
\end{equation}
where $d\mathbf{l}:=\frac{\partial\mathbf{x(}t,\ell\mathbf{)}}{\partial\ell
}d\ell$ and Faraday's law reads:
\begin{equation}%
{\displaystyle\int\nolimits_{\mathbf{\Gamma}_{t}}}
(\mathbf{E+v\times B)}\cdot d\mathbf{l}=\mathcal{-}\frac{d}{dt}%
{\displaystyle\int\nolimits_{\mathcal{S}_{t}}}
\mathbf{B}\cdot\mathbf{n}\text{ }da.\label{03}%
\end{equation}
We want to prove that Eq.(\ref{03}) is equivalent to
\begin{equation}%
{\displaystyle\int\nolimits_{\mathbf{\Gamma}_{t}}}
\mathbf{E}\cdot d\mathbf{l}=-%
{\displaystyle\int\nolimits_{\mathcal{S}_{t}}}
\frac{\partial\mathbf{B}}{\partial t}\cdot\mathbf{n}\text{ }da,\label{04}%
\end{equation}
from where it trivially follows the differential form of Faraday's law, i.e.,
\begin{equation}
\mathbf{\nabla}\times\mathbf{E}+\frac{\partial\mathbf{B}}{\partial
t}=0.\label{faraday law}%
\end{equation}

Those statements will be proved in Section 3, but first we shall need to
recall a few mathematical results concerning differentiable vector fields, in
Section 2.

\section{Some Identities Involving the Integration of Differentiable Vector
Fields}

Let $U\subset\mathbb{R}^{3}$ be a convex and simply-connected (open) region,
\ $\mathbf{X}:\mathbb{R\times}U\rightarrow\mathbb{R}^{3}$%
,$(t,\mathbf{x)\mapsto X(}t,\mathbf{x})$ be a generic differentiable vector
field and let $\mathbf{v}:\mathbb{R\times}U\rightarrow\mathbb{R}^{3}$ be a
differentiable \textit{velocity vector field} of a \textit{fluid flow}. A
\textit{trajetory} (of a \textquotedblleft fluid particle\textquotedblright%
)\footnote{An integral line (or a stream line)\ of $\mathbf{v}$\textbf{ }at
fixed $t$ is a mapping $\phi_{\mathbf{x}}:\mathbb{R\rightarrow R}^{3}$,
$s\mapsto\phi_{\mathbf{x}}(s)$, $\phi(0)=\mathbf{x}$, such that $\frac{d}%
{ds}\phi_{\mathbf{x}}(s)=\mathbf{v(}t,\phi_{\mathbf{x}}(s)\mathbf{)}$. Note
that if $\mathbf{v}$ is stationary ($\frac{\partial\mathbf{v}}{\partial t}%
=0$), then the concepts of integral lines and trajectories coincide
\cite{chorin}.} associated to $\mathbf{v}$ passing through a given
$\mathbf{x}\in\mathbb{R}^{3}$ is a smooth curve $\mathbf{\sigma}_{\mathbf{x}%
}:\mathbb{R\rightarrow}\mathbb{R}^{3}$, $t\rightarrow\mathbf{\sigma
}_{\mathbf{x}}(t)=\mathbf{\sigma}(t,\mathbf{x)}$ which at $t=0$ is at
$\mathbf{x}$ (i.e., $\mathbf{\sigma}_{\mathbf{x}}(0)\mathbf{=x}$ ) and such
that its tangent vector at $\mathbf{\sigma}(t,\mathbf{x)}$ is
\begin{equation}
\frac{\partial}{\partial t}\mathbf{\sigma}(t,\mathbf{x)}=\mathbf{v(}%
t,\mathbf{\sigma}(t,\mathbf{x))}.\label{tangent}%
\end{equation}
Let moreover $\mathbf{\sigma}_{t}:U\rightarrow\mathbb{R}^{3},\mathbf{\sigma
}_{t}(\mathbf{x})=\mathbf{\sigma}(t,\mathbf{x)}$. We call $\mathbf{\sigma}%
_{t}$ the \textit{fluid flow map}. Let $J=(0,1)\in\mathbb{R}$ and let
$\mathbf{\Gamma}$ be a closed loop parametrized by $\mathbf{\Gamma
}:J\rightarrow\mathbb{R}^{3},$ $\ell\mapsto$ $\mathbf{\Gamma}(\ell
):=\mathbf{x(\ell)}$ and denote by $\Gamma_{t}=\mathbf{\sigma}_{t}(\Gamma)$
the loop \textit{transported} by the flow. Then%
\begin{equation}
\mathbf{\sigma}(t,\mathbf{x(\ell)):=x}(t,\ell)\label{looptransp}%
\end{equation}
is clearly a parametrization of $\mathbf{\Gamma}_{t}$. We have the
\ proposition:\newpage

\textbf{Proposition}%

\begin{subequations}
\begin{align}
\frac{d}{dt}\int\nolimits_{\Gamma_{t}}\mathbf{X}\cdot d\mathbf{l} &
=\int\nolimits_{\Gamma_{t}}\frac{D}{Dt}\mathbf{X}\cdot d\mathbf{l}%
+\int\nolimits_{\Gamma_{t}}\mathbf{X\cdot\lbrack(}d\mathbf{l}\cdot
\mathbf{\nabla)v],}\label{1a}\\
&  =\int\nolimits_{\Gamma_{t}}\frac{D}{Dt}\mathbf{X}\cdot d\mathbf{l}%
+\int\nolimits_{\Gamma_{t}}[\mathbf{X\times(\nabla}\times\mathbf{v)}]\cdot
d\mathbf{l}+\int\nolimits_{\Gamma_{t}}[(\mathbf{X}\cdot\mathbf{\nabla
})\mathbf{v})]\cdot d\mathbf{l}\label{1b}\\
&  =\int\nolimits_{\Gamma_{t}}\frac{\partial}{\partial t}\mathbf{X}\cdot
d\mathbf{l}-\int\nolimits_{\Gamma_{t}}[\mathbf{v}\times(\mathbf{\nabla\times
X)}]\cdot d\mathbf{l},\label{1c}%
\end{align}
where%
\end{subequations}
\begin{equation}
\frac{D}{Dt}\mathbf{X:=}\frac{\partial}{\partial t}\mathbf{X+(v\cdot\nabla
})\mathbf{X}\label{matderiv}%
\end{equation}
is the so-called \textit{material derivative\footnote{Mind that the
\textit{material derivative }is a derivative taken along a path
$\mathbf{\sigma}_{t}$ with tangent vector $\left.  \mathbf{v}\right\vert
_{\mathbf{\sigma}_{x}}$. It is frequently used in fluid mechanics, where it
describes the total time rate of change of a given quantity as viewed by a
fluid particle moving on $\mathbf{\sigma}_{x}$.}} and
\begin{equation}
d\mathbf{l}=\frac{\partial}{\partial\ell}\mathbf{\sigma}(t,\mathbf{x(\ell
))}d\mathbf{\ell=}\frac{\partial\mathbf{x(}t\mathbf{,\ell)}}{\partial\ell
}d\mathbf{\ell}\label{dl}%
\end{equation}
is the tangent line element\footnote{Take notice that $d\mathbf{l}$ is not an
explicit function of the cartesian coordinates $(x,y,z)$.} of $\Gamma_{t}$ at
\textbf{$\sigma$}$(t,\mathbf{x(\ell))}$.

\begin{proof}
We can write%
\begin{align}
\frac{d}{dt}\int\nolimits_{\Gamma_{t}}\mathbf{X}\cdot d\mathbf{l} &  =\frac
{d}{dt}\int\limits_{0}^{1}\mathbf{X(}t,\mathbf{\sigma}(t,\mathbf{x(\ell
)))}\cdot\frac{\partial}{\partial\ell}\mathbf{\sigma}(t,\mathbf{x(\ell
))}d\mathbf{\ell}\nonumber\\
&  =\int\limits_{0}^{1}\frac{d}{dt}\left[  \mathbf{X(}t,\mathbf{\sigma
}(t,\mathbf{x(\ell)))}\right]  \cdot\frac{\partial}{\partial\ell
}\mathbf{\sigma}(t,\mathbf{x(\ell))}d\mathbf{\ell}\nonumber\\
&  +\int\limits_{0}^{1}\mathbf{X(}t,\mathbf{\sigma}(t,\mathbf{x(\ell)))}%
\cdot\frac{\partial}{\partial t}\frac{\partial}{\partial\ell}\mathbf{\sigma
}(t,\mathbf{x(\ell))}d\mathbf{\ell.}\label{2}%
\end{align}
\ Now, taking into account that for each $\mathbf{x}(\ell)$, $\frac{\partial
}{\partial t}\mathbf{\sigma}(t,\mathbf{x)=v(}t\mathbf{,\sigma}%
(t,\mathbf{x(\ell)))}$ we have%
\begin{equation}
\frac{D}{Dt}\left[  \mathbf{X(}t,\mathbf{\sigma}(t,\mathbf{x(\ell)))}\right]
=\frac{\partial}{\partial t}\mathbf{X\mathbf{(}}t\mathbf{,\mathbf{\sigma}%
(}t\mathbf{,\mathbf{x(\ell)))}+(v}\cdot\mathbf{\nabla})\mathbf{X(}%
t,\mathbf{\sigma}(t,\mathbf{x(\ell))),}\label{2'}%
\end{equation}
\ hence, the first term in the right side of Eq.(\ref{2}) can be written as%
\begin{equation}
\int\limits_{0}^{1}\frac{d}{dt}\left[  \mathbf{X(}t,\mathbf{\sigma
}(t,\mathbf{x(\ell)))}\right]  \cdot\frac{\partial}{\partial\ell
}\mathbf{\sigma}(t,\mathbf{x(\ell))}d\mathbf{\ell}=\int\nolimits_{\Gamma_{t}%
}\mathbf{[}\frac{\partial}{\partial t}\mathbf{X+(v}\cdot\mathbf{\nabla
})\mathbf{X]\cdot}d\mathbf{l}=\int\nolimits_{\Gamma_{t}}\frac{D}{Dt}%
\mathbf{X}\cdot d\mathbf{l}.\label{0}%
\end{equation}

Also writing $\mathbf{\sigma}(t,\mathbf{x(\ell))=(}x^{1}(t,\ell),x^{2}%
(t,\ell),x^{3}(t,\ell))$ we see that the last term in Eq.(\ref{2}) can be
written as:%
\begin{align}
\int\limits_{0}^{1}\mathbf{X(}t,\mathbf{\sigma}(t,\mathbf{x(\ell)))}\cdot
\frac{\partial}{\partial t}\frac{\partial}{\partial\ell}\mathbf{\sigma
}(t,\mathbf{x(\ell))}d\mathbf{\ell} &  =\int\limits_{0}^{1}\mathbf{X(}%
t,\mathbf{\sigma}(t,\mathbf{x(\ell)))\cdot}\left[  \frac{\partial}%
{\partial\ell}\mathbf{v}(t,\mathbf{\sigma}(t,\mathbf{x(\ell)))}d\mathbf{\ell
}\right]  \nonumber\\
&  =\int\nolimits_{\Gamma_{t}}\mathbf{X\cdot\lbrack(}d\mathbf{l}%
\cdot\mathbf{\nabla)v].}\label{3}%
\end{align}
We now recall that for arbitrary differentiable vector fields $\mathbf{a,b}%
:U\rightarrow\mathbb{R}^{3}$ it holds%
\begin{equation}
\mathbf{\nabla}(\mathbf{a\cdot b})=(\mathbf{a\cdot\nabla})\mathbf{b}%
+(\mathbf{b\cdot\nabla})\mathbf{a}+\mathbf{a}\times(\mathbf{\nabla\times
b})+\mathbf{b}\times(\mathbf{\nabla}\times\mathbf{a}).\label{4}%
\end{equation}
Setting $\mathbf{a=}$ $d\mathbf{l}$ and $\mathbf{b=}$ $\mathbf{v}$ and noting
that $(\mathbf{v\cdot\nabla})d\mathbf{l=v\times(}$\textbf{$\nabla$}$\times
d\mathbf{\mathbf{l})=0,}$ it implies that%
\begin{equation}
\mathbf{(}d\mathbf{l}\cdot\mathbf{\nabla)v=\nabla}(d\mathbf{l}\cdot
\mathbf{v)-}d\mathbf{l}\times(\mathbf{\nabla}\times\mathbf{v}).\label{5}%
\end{equation}
We need also to recall the well known identity%
\begin{equation}
\mathbf{a\cdot(b}\times\mathbf{c)=b\cdot(c}\times\mathbf{a)}\label{5-b}%
\end{equation}
\newline which implies setting $\mathbf{a=X,}$ $\mathbf{b=}$ $d\mathbf{l}$ and
$\mathbf{c=}$ $(\mathbf{\nabla}\times\mathbf{v}),$ that%
\begin{align}
-\mathbf{X\cdot\lbrack}d\mathbf{l}\times(\mathbf{\nabla}\times\mathbf{v})]  &
=-d\mathbf{l}\cdot\lbrack(\mathbf{\nabla}\times\mathbf{v)\times X}%
],\nonumber\\
& =d\mathbf{l}\cdot\lbrack\mathbf{X}\times(\mathbf{\nabla}\times
\mathbf{v)}],\label{id}%
\end{align}
and also the \textit{not }so well known identity\footnote{See the Appendix for
a proof of this identity.}%
\begin{equation}
\mathbf{X\cdot\lbrack\nabla(}d\mathbf{l\cdot\mathbf{v})]=[(X}\cdot
\mathbf{\nabla})\mathbf{v}]\cdot d\mathbf{l},\label{6}%
\end{equation}
to write that%
\begin{align}
\int\nolimits_{\Gamma_{t}}\mathbf{X\cdot\lbrack(}d\mathbf{l}\cdot
\mathbf{\nabla)v]} &  \mathbf{=}-\int\nolimits_{\Gamma_{t}}\mathbf{X\cdot
\lbrack}d\mathbf{l\times(\nabla\times\mathbf{v})]+}\int\nolimits_{\Gamma_{t}%
}\mathbf{X\cdot\lbrack\nabla(}d\mathbf{l\cdot\mathbf{v})]}\nonumber\\
&  =\int\nolimits_{\Gamma_{t}}[\mathbf{X\times(\nabla}\times\mathbf{v)}]\cdot
d\mathbf{l}+\int\nolimits_{\Gamma_{t}}[(\mathbf{X}\cdot\mathbf{\nabla
})\mathbf{v})]\cdot d\mathbf{l}.\label{7}%
\end{align}
Finally, using Eq.(\ref{0}) and Eq.(\ref{7}) completes the proof of
Eq.(\ref{1a}) and Eq.(\ref{1b}). Also, from Eq.(\ref{1b}) if we use
Eq.(\ref{4}), setting $\mathbf{a}=\mathbf{X}$ and $\mathbf{b}=\mathbf{v}$ and
noting that $\int\nolimits_{\Gamma_{t}}[\mathbf{\nabla(X}\cdot\mathbf{v}%
)]\cdot d\mathbf{l}=0$, it follows%
\begin{align*}
&  \frac{d}{dt}\int\nolimits_{\Gamma_{t}}\mathbf{X}\cdot d\mathbf{l}\\
&  =\int\nolimits_{\Gamma_{t}}\frac{\partial}{\partial t}\mathbf{X}\cdot
d\mathbf{l}+\int\nolimits_{\Gamma}[(\mathbf{v}\cdot\mathbf{\nabla}%
)\mathbf{X})]\cdot d\mathbf{l}+\int\nolimits_{\Gamma_{t}}[\mathbf{X\times
(\nabla}\times\mathbf{v)}]\cdot d\mathbf{l}+\int\nolimits_{\Gamma}%
[(\mathbf{X}\cdot\mathbf{\nabla})\mathbf{v})]\cdot d\mathbf{l}\\
&  =\int\nolimits_{\Gamma_{t}}\frac{\partial}{\partial t}\mathbf{X}\cdot
d\mathbf{l}-\int\nolimits_{\Gamma_{t}}[\mathbf{v\times}(\mathbf{\nabla\times
X)}]\cdot d\mathbf{l}.
\end{align*}
from where the proof of Eq.(\ref{1c}) follows immediately.
\end{proof}

\begin{remark}
Before proceeding, we recall that if $\mathbf{X=v\ }$we have
\begin{equation}
\frac{d}{dt}\int\nolimits_{\Gamma_{t}}\mathbf{v}\cdot d\mathbf{l}%
=\int\nolimits_{\Gamma_{t}}\frac{D}{Dt}\mathbf{v}\cdot d\mathbf{l}%
,\label{kelvin}%
\end{equation}
a result that is known in fluid mechanics as \textit{Kelvin's circulation
theorem} \emph{(see, e.g., \cite{chorin,saffman})}.
\end{remark}

Now,%
\begin{equation}
\frac{d}{dt}\int\nolimits_{\Gamma_{t}}\mathbf{X}\cdot d\mathbf{l}=\frac{d}%
{dt}\int\nolimits_{\mathcal{S}_{t}}(\mathbf{\nabla}\times\mathbf{X)\cdot
n}\text{ }da,\label{8}%
\end{equation}
where, if $\mathcal{S}$ is a smooth surface such that $\partial\mathcal{S}%
=\Gamma$, then $\mathcal{S}_{t}=\mathbf{\sigma}_{t}(\mathcal{S})$. Also
$\mathbf{n}$ is the normal vector field to $S_{t}$. Then using Eq.(\ref{1c})
we can write:%

\begin{align}
\frac{d}{dt}\int\nolimits_{\mathcal{S}_{t}}(\mathbf{\nabla}\times
\mathbf{X)\cdot n}\text{ }da &  =\int\nolimits_{\Gamma_{t}}\frac{\partial
}{\partial t}\mathbf{X}\cdot d\mathbf{l}-\int\nolimits_{\Gamma_{t}%
}[\mathbf{v\times}(\mathbf{\nabla\times X)}]\cdot d\mathbf{l}\nonumber\\
&  =\int\nolimits_{\mathcal{S}_{t}}\frac{\partial}{\partial t}(\mathbf{\nabla
}\times\mathbf{X)\cdot n}\text{ }da-\int\nolimits_{\mathcal{S}_{t}%
}\mathbf{\nabla}\times\lbrack\mathbf{v\times}(\mathbf{\nabla\times X)}%
]\cdot\mathbf{n}\text{ }da.\label{9}%
\end{align}
Also, denoting $\mathbf{Y:}=\mathbf{\nabla\times X}$ we can write%
\begin{equation}
\frac{d}{dt}\int\nolimits_{\mathcal{S}_{t}}\mathbf{Y\cdot n}\text{ }%
da=\int\nolimits_{\mathcal{S}_{t}}\left[  \frac{\partial}{\partial
t}\mathbf{Y}-\mathbf{\nabla}\times(\mathbf{v\times Y)}\right]  \cdot
\mathbf{n}\text{ }da.\label{10}%
\end{equation}
Despite Eq.(\ref{10}), for a general differentiable vector field
$\mathbf{Z}:\mathbb{R\times}U\rightarrow\mathbb{R}^{3}$ such that
$\mathbf{\nabla}\cdot\mathbf{Z}\neq0$ we have
\begin{equation}
\frac{d}{dt}\int\nolimits_{\mathcal{S}_{t}}\mathbf{Z\cdot n}\text{ }%
da=\int\nolimits_{\mathcal{S}_{t}}\left[  \frac{\partial}{\partial
t}\mathbf{Z}+\mathbf{v(\nabla}\cdot\mathbf{Z)}-\mathbf{\nabla}\times
(\mathbf{v\times Z)}\right]  \cdot\mathbf{n}\text{ }da,\label{11}%
\end{equation}
the so-called \textit{Helmholtz identity }\cite{helmholtz}\textit{. }Note that
the identity is also mentioned in \cite{hertz}. A proof of Helmholtz identity
can be obtained using arguments similar to the ones used in the proof of
Eq.(\ref{1a}). Some textbooks quoting Helmholtz identity are
\cite{abraham,jackson,panofski,sommerfeld,whites}. However, we emphasize that
the proof of Faraday's law of induction presented in all the textbooks just
quoted are always for very particular situations and definitively not
satisfactory from a mathematical point of view.

We now want to use the above results to prove Eq.(\ref{03}) and Eq.(\ref{04}).

\section{Proofs of Eq.(\ref{03}) and Eq.(\ref{04})}

We start remembering that in Maxwell theory we have that the $\mathbf{E}%
$\textbf{ }and $\mathbf{B}$ fields are derived from potentials, i.e.,
\begin{align}
\mathbf{E} &  =-\mathbf{\nabla}\phi-\frac{\partial\mathbf{A}}{\partial
t},\nonumber\\
\mathbf{B} &  =\mathbf{\nabla}\times\mathbf{A,}\label{00}%
\end{align}
where $\phi:\mathbb{R}\times\mathbb{R}^{3}\rightarrow\mathbb{R}$ is the scalar
potential and $\mathbf{A}:\mathbb{R}\times\mathbb{R}^{3}\mapsto\mathbb{R}$ is
the (magnetic) vector potential. If Eq.(\ref{00}) is taken into account we can
immediately derive Eq.(\ref{03}). All we need is to use the results just
derived in Section 2 taking $\mathbf{X}=\mathbf{A}$. Indeed, the first line of
Eq.(\ref{9}) then becomes%
\[
\frac{d}{dt}\int\nolimits_{\mathcal{S}_{t}}(\mathbf{\nabla}\times
\mathbf{A)\cdot n}\text{ }da=\int\nolimits_{\mathbf{\Gamma}_{t}}\frac
{\partial}{\partial t}\mathbf{A}\cdot d\mathbf{l}-\int
\nolimits_{\mathbf{\Gamma}_{t}}[\mathbf{v\times}(\mathbf{\nabla\times
A)}]\cdot d\mathbf{l},
\]
or%
\begin{align}
\frac{d}{dt}\int\nolimits_{\mathcal{S}_{t}}\mathbf{B\cdot n}\text{ }da &
=\int\nolimits_{\mathbf{\Gamma}_{t}}\frac{\partial}{\partial t}\mathbf{A}\cdot
d\mathbf{l}-\int\nolimits_{\mathbf{\Gamma}_{t}}(\mathbf{v\times B}]\cdot
d\mathbf{l}\nonumber\\
&  =\int\nolimits_{\mathbf{\Gamma}_{t}}\left(  \frac{\partial}{\partial
t}\mathbf{A+\nabla}\phi-\mathbf{v\times B}\right)  \cdot d\mathbf{l}%
\nonumber\\
&  =-\int\nolimits_{\mathbf{\Gamma}_{t}}(\mathbf{E+v\times B)}\cdot
d\mathbf{l.}\label{f1}%
\end{align}
To obtain Eq.(\ref{04}) we recall that from the second line of Eq.(\ref{9}) we
can write (using Stokes theorem)%
\begin{align}
\frac{d}{dt}\int\nolimits_{\mathcal{S}_{t}}\mathbf{B\cdot n}\text{ }da &
=\int\nolimits_{\mathcal{S}_{t}}\frac{\partial}{\partial t}\mathbf{B\cdot
n}\text{ }da-\int\nolimits_{\mathcal{S}_{t}}\mathbf{\nabla}\times
\lbrack\mathbf{v\times B}]\cdot\mathbf{n}\text{ }da\nonumber\\
&  =\int\nolimits_{\mathcal{S}_{t}}\frac{\partial}{\partial t}\mathbf{B\cdot
n}\text{ }da-\int\nolimits_{\mathbf{\Gamma}_{t}}(\mathbf{v\times B})\cdot
d\mathbf{l}.\label{f2}%
\end{align}
Comparing the second member of Eq.(\ref{f1}) and \ Eq.(\ref{f2}) we get
Eq.(\ref{04}), i.e.,%
\begin{equation}
\int\nolimits_{\mathbf{\Gamma}_{t}}\mathbf{E}\cdot d\mathbf{l}=-\int
\nolimits_{\mathcal{S}_{t}}\frac{\partial}{\partial t}\mathbf{B\cdot n}\text{
}da,\label{F3}%
\end{equation}
from where the differential form of Faraday's law follows.

\begin{remark}
We end this section by recalling that in the physical world the \textit{real}
circuits are not filamentary and worse, are not described by smooth closed
curves. However, if the closed curve representing a `filamentary circuit' is
made of finite number of sections that are smooth, we can yet apply the above
formulas with the integrals meaning Lebesgue integrals.
\end{remark}

\section{Conclusions}

Recently a paper \cite{red1} \ titled \ `Faraday's Law via the Magnetic Vector
Potential', has been commented in \cite{khol} and replied in \cite{red2}.
Thus, the author of \cite{red1}, claims to have presented an \textquotedblleft
alternative\textquotedblright\ derivation for Faraday's law for a filamentary
circuit which is moving with an arbitrary velocity and which is changing its
shape, using directly the vector potential $\mathbf{A}$ instead of the
magnetic field $\mathbf{B}$ and the electric field $\mathbf{E}$ (which is the
one presented in almost all textbooks).

Now, \cite{khol} correctly identified that the derivation in \cite{red1} is
wrong, and that author agreed with that in \cite{red2}. Here we want to recall
that a presentation of Faraday's law in terms of the magnetic vector potential
$\mathbf{A}$ already appeared in Maxwell treatise \cite{maxwell}, using big
formulas involving the components of the vector fields involved. We recall
also that a formulation \ of Faraday's law in terms of $\mathbf{A}$ using
modern vector calculus has been given by Gamo more than 30 years ago
\cite{gamo}. In Gamo's paper (not quoted in \cite{khol,red1,red2})
Eqs.(\ref{1c}) appear for the special case in which $\mathbf{X}=\mathbf{A}$
(the vector potential) and \ $\mathbf{B=}$\textbf{$\nabla$}$\mathbf{\times A}$
(the magnetic field), i.e.,%

\begin{equation}
\frac{d}{dt}\int\nolimits_{\mathbf{\Gamma}_{t}}\mathbf{A}\cdot d\mathbf{l}%
=\int\nolimits_{\mathbf{\Gamma}_{t}}\frac{\partial}{\partial t}\mathbf{A}\cdot
d\mathbf{l}-\int\nolimits_{\mathbf{\Gamma}_{t}}[\mathbf{v}\times
(\mathbf{\nabla\times A)}]\cdot d\mathbf{l}.\label{12'}%
\end{equation}

Thus, Eq.(\ref{12'}) also appears in \cite{red1} (it is there Eq.(9)).
However, in footnote 3 of \cite{red1} it is said that Eq.(\ref{12'}) is
equivalent to \textquotedblleft$\frac{d}{dt}\int\nolimits_{\Gamma_{t}%
}\mathbf{A}\cdot d\mathbf{l}=\int\nolimits_{\Gamma_{t}}\frac{D}{Dt}%
\mathbf{A}\cdot d\mathbf{l}$ \textquotedblright, where the term $\int
\nolimits_{\Gamma}[(\mathbf{A}\cdot\mathbf{\nabla})\mathbf{v})]\cdot
d\mathbf{l}$ is missing. This is the error that has been observed by authors
\cite{khol}, which also presented a proof of Eq.(\ref{1b}), which however is
not very satisfactory from a mathematical point of view, that being one of the
reasons why we decided to write this note presenting a correct derivation of
Faraday's law in terms of $\mathbf{A}$ and its relation with Helmholtz
formula. Another reason is that there are still people (e.g., \cite{phipps})
that do not understand that Eq.(\ref{03}) and Eq.(\ref{04}) are equivalent and
think that Eq.(\ref{03}) implies the form of Maxwell equations as given by
Hertz, something that we know since a long time that is wrong \cite{miller}.

We also want to observe that Jackson's proof of Faraday's law using
\ `Galilean invariance' is \ valid only for a \ filamentary circuit moving
without deformation with a constant velocity. The proof we presented is
general and valid in Special Relativity, since it is based on trustful
mathematical identities and in the Lorentz force law applied in the laboratory
frame with the motion and deformation of the filamentary circuit
mathematically well described.

\appendix{}

\section{Proof of the Identity in Eq.(\ref{6})}

We know from Eq.(\ref{5}) that%
\begin{equation}
\mathbf{\nabla}(d\mathbf{l}\cdot\mathbf{v)=\mathbf{(}}d\mathbf{\mathbf{l}%
\cdot\nabla\mathbf{)v}+}d\mathbf{l}\times(\mathbf{\nabla}\times\mathbf{v}%
).\label{App-A0}%
\end{equation}
Let $\left\{  e^{1},e^{2},e^{3}\right\}  $ be an orthonormal base of
$\mathbb{R}^{3}$. We can write, using \textit{Einstein convention,}%
\begin{equation}
\mathbf{(\nabla\times\mathbf{v})}=e^{i}\mathbf{\partial}_{i}\times
\mathbf{v}=e^{i}\times\partial_{i}\mathbf{v,}\label{App=A1}%
\end{equation}
where $\mathbf{\nabla}=\left(  \partial_{1},\partial_{2},\partial_{3}\right)
=$ $e^{1}\frac{\partial}{\partial x^{1}}+e^{2}\frac{\partial}{\partial x^{2}%
}+e^{3}\frac{\partial}{\partial x^{3}}=e^{i}\mathbf{\partial}_{i},$ with
$\partial_{i}=\frac{\partial}{\partial x^{i}}$ and $\{x^{i}\}$, $i=1,2,3$ are
Cartesian coordinates. It follows then%
\begin{equation}
d\mathbf{l}\times\mathbf{(\nabla\times\mathbf{v})}=d\mathbf{l}\times
(e^{i}\times\partial_{i}\mathbf{v).}\label{App-A2}%
\end{equation}
Using the known identity $\mathbf{a}\times\mathbf{b}\times\mathbf{c}%
=(\mathbf{a\cdot c})\mathbf{b}-(\mathbf{a\cdot b})\mathbf{c}$ in
Eq.(\ref{App-A2}), we obtain%
\begin{equation}
d\mathbf{l}\times(e^{i}\times\partial_{i}\mathbf{v)}=(d\mathbf{l\cdot\partial
}_{i}\mathbf{v})e^{i}-(d\mathbf{l}\cdot e^{i})\partial_{i}\mathbf{v.}%
\label{App-A3}%
\end{equation}
On the other hand, considering $d\mathbf{l}=\left(  dl_{1},dl_{2}%
,dl_{3}\right)  =dl_{i}e^{i}$, we have%
\begin{equation}
\mathbf{\mathbf{(}}d\mathbf{\mathbf{l}\cdot\nabla\mathbf{)v=(}}dl_{i}%
\partial_{i})\mathbf{v=(}d\mathbf{l}\cdot e^{i})\partial_{i}\mathbf{v.}%
\label{App-A4}%
\end{equation}
Hence, substituting Eq.(\ref{App-A3}) and Eq.(\ref{App-A4}) in
Eq.(\ref{App-A0}), we can rewrite it as%
\begin{align}
\mathbf{\nabla}(d\mathbf{l}\cdot\mathbf{v)} &  \mathbf{=}\mathbf{(}%
d\mathbf{l}\cdot e^{i})\partial_{i}\mathbf{v+}(d\mathbf{l\cdot\partial}%
_{i}\mathbf{v})e^{i}-(d\mathbf{l}\cdot e^{i})\partial_{i}\mathbf{v}\nonumber\\
&  =(d\mathbf{l\cdot\partial}_{i}\mathbf{v})e^{i}.\label{App-A5}%
\end{align}
From this last result, it is easy to see that%
\begin{align*}
\mathbf{X}\cdot\left[  \mathbf{\nabla}\left(  d\mathbf{l\cdot v}\right)
\right]   &  =\mathbf{X}\cdot\left[  (d\mathbf{l\cdot\partial}_{i}%
\mathbf{v})e^{i}\right]  =X^{i}(d\mathbf{l\cdot\partial}_{i})\mathbf{v=}%
d\mathbf{l\cdot(}X^{i}\partial_{i}\mathbf{)v}\\
&  =d\mathbf{l}\cdot\left[  \mathbf{(X\cdot\nabla)v}\right]  =\left[
\mathbf{(X\cdot\nabla)v}\right]  \cdot d\mathbf{l,}%
\end{align*}
where $\mathbf{X=(}X^{1},X^{2},X^{3})=X^{i}e_{i}$, $e_{i}\cdot e^{j}%
=\delta_{i}^{j}$.


\begin{thebibliography}{99}                                                                                               %


\bibitem {abraham}Abraham, M. and Becker, R., \textit{The Classical Theory of
Electricity and Magnetism}, Blackie, London, 1932.

\bibitem {chorin}Chorin, A. J. and Mardsen, J. E., \textit{A Mathematical
Introduction to Fluid Mechanics} (third edition), Springer-Verlag, New York, 1993.

\bibitem {gamo}Gamo, S. A., General Formulation of Faraday's Law of Induction,
\textit{Proc. IEEE} \textbf{67}, 676-677\textbf{ }(1979).

\bibitem {helmholtz}Helmholtz, H.,\textit{ Gesammelt Scriften} vol. I (3), pp.
597-603, Olms-Weidmann, Hildeshein 2003, (a reprint of
\textit{Wissenschaftliche Abhandlungen} vol. \textbf{3}, Johann Ambrosius
Barth, Leipzig, 1895).

\bibitem {hertz}Hertz, H. R., \textit{Electric Waves}, MacMillan, London, pp.
243-247, 1893 (also, Dover, New York, 1962).

\bibitem {jackson}Jackson, J. D., \textit{Classical Electrodynamics }(third
edition) J., Wiley \& Sons, Inc., New York, 1999.

\bibitem {khol}Kholmetskii, A. K., Missevitch, O., and \ Yarman, T., Comment
on the Note: `Faraday's Law via the Magnetic Vector Potential' by Dragan V.
Red\v{z}i\'{c}, \textit{Eur. J. Phys.} \textbf{29},\textbf{ }L1-L4 (2008).

\bibitem {maxwell}Maxwell, J. C., \textit{A Treatise of Electricity and
Magnetism}, pp. 238-243, Dover, New York, 1954.

\bibitem {miller}Miller, A. I.,\textit{ Albert Einstein's Special Theory of
Relativity}, pp. 16, Addison-Wesley Publ. Co., Inc., Reading, Ma., 1981.

\bibitem {panofski}Panofski, W. K. H. and Phillips, M., \textit{Classical
Electricity and Magnetism}, pp. 160-162, Addison-Wesley Publ. Co., Reading MA, 1962.

\bibitem {phipps}Phipps, T. E. Jr., \textit{Old Physics for New - A Worldview
Alternative to Einstein's Relativity Theory}, Apeiron, Montreal, 2006.

\bibitem {red1}Red\v{z}i\'{c}, D. V., Faraday's Law via the Magnetic Vector
Potential, \textit{Eur. J. Phys.,} \textbf{28},\textbf{ }N7-N10, (2007).

\bibitem {red2}Red\v{z}i\'{c}, D. V., Reply to\ `Comment on \ `Faraday's Law
via the Magnetic Vector Potential' ', \textit{Eur. J. Phys}., \textbf{29, }L5, (2008).

\bibitem {rodcap2007}Rodrigues, W. A. Jr. and Capelas de Oliveira, E.,
\textit{The Many Faces of Maxwell, Dirac and Einstein Equations. A Clifford
Bundle Approach}, Lecture Notes in Physics \textbf{722}, Springer, Heidelberg, 2007.

\bibitem {sawu}Sachs, R. K., and Wu, H., \textit{General Relativity for
Mathematicians}, Springer-Verlag, New York, 1977.

\bibitem {saffman}Saffman, P.G., \textit{Vortex Dynamics}, Cambridge
Monographs on Mechanics and Applied Mathematics, Cambridge Univ. Press,
Cambridge, 1992.

\bibitem {sommerfeld}Sommerfeld, A., \textit{Electrodynamics}, pp. 285-287,
Academic Press, New York, 1952.

\bibitem {whites}Paul, C. R., Whites, K. W. and Nasar, A. A.,
\textit{Introduction to Electromagnetic Fields} (third edition), pp. 653-658 ,
McGraw-Hill, Boston, 1998.
\end{thebibliography}
\end{document}